\documentclass[12pt,letterpaper]{article}
\pdfoutput=1
\usepackage{jheppub}
\usepackage{amssymb}
\usepackage{color}
\usepackage{slashed}
\usepackage{amsmath}
\usepackage{amsfonts}
\usepackage{amssymb}
\usepackage{graphicx,subcaption}
\usepackage{caption}
\usepackage{subcaption}
\usepackage{float}
\usepackage{tikz}
\usetikzlibrary[arrows]
\usepackage{amsmath}
\usepackage{epsfig}
\usepackage{appendix}
\usepackage{listings}
\usepackage{xcolor}
\usepackage{mathabx}

\def\bR{{\mathbb R}}

\newcommand{\be}{\begin{equation}}
\newcommand{\ee}{\end{equation}}
\newcommand{\vev}[1]{{\left< {#1} \right>}}
\newcommand{\tlambda}{\tilde \lambda}

\title{The planar limit of ${\cal N}=2$ chiral correlators}

\author{Bartomeu Fiol}
\author{and Alan Rios Fukelman}

\affiliation{Departament de F{\'\i}sica Qu\`antica i Astrof\'isica i \\Institut de Ci{\`e}ncies del Cosmos, 
Universitat de Barcelona,
Mart{\'\i}\ i Franqu{\`e}s 1, 08028 Barcelona, Catalonia, Spain}

\emailAdd{bfiol@ub.edu}
\emailAdd{ariosfukelman@icc.ub.edu}

\abstract{We derive the planar limit of 2- and 3-point functions of single-trace chiral primary operators of ${\cal N}=2$ SQCD on $S^4$, to all orders in the 't Hooft coupling. In order to do so, we first obtain a combinatorial expression for the planar free energy of a hermitian matrix model with an infinite number of arbitrary single and double trace terms in the potential; this solution might have applications in many other contexts. We then use these results to evaluate the analogous planar correlation functions on $\bR^4$. Specifically, we compute all the terms with a single value of the $\zeta$ function for a few planar 2- and 3-point functions, and conjecture general formulas for these terms for all 2- and 3-point functions on $\bR^4$.
}

\begin{document}
\maketitle
\section{Introduction}

Correlation functions of local operators are among the most basic quantities of interest in any quantum field theory, yet in most cases their evaluation is prohibitively difficult. This state of affairs can improve in theories with additional symmetries, like conformal invariance and/or supersymmetry. In this work we are going to study a particular instance of such tractable correlation functions, the so-called extremal n-point functions of chiral primary operators (CPOs) of four dimensional Lagrangian ${\cal N}=2$ superconformal field theories (SCFTs) \cite{Papadodimas:2009eu}. For these correlation functions, the coupling and spacetime dependences factorize, and the spacetime dependence is completely fixed, thus reducing the problem to the - still very difficult - determination of the dependence on the marginal coupling. 

In recent years, the study of these correlation functions has been approached from different angles, often in combination. A first approach \cite{Baggio:2014sna, Baggio:2014ioa, Baggio:2015vxa, Baggio:2016skg} uses a 4d analog of the $tt^*$ equations \cite{Papadodimas:2009eu}. More recently, it has been shown  \cite{Gerchkovitz:2014gta, Gerchkovitz:2016gxx} that the evaluation of closely related n-point functions on $S^4$ can be reduced through supersymmetric localization to matrix model computations; in turn, a Gram-Schmidt orthogonalization procedure applied to these $S^4$ correlators yields the correlators on $\bR^4$ \cite{Rodriguez-Gomez:2016ijh, Rodriguez-Gomez:2016cem, Billo:2017glv, Beccaria:2020hgy,Galvagno:2020cgq, Beccaria:2021hvt}. Alternatively, the large R-charge limit \cite{Hellerman:2015nra} of these correlation functions has been studied in \cite{Hellerman:2017sur, Bourget:2018obm, Grassi:2019txd, Beccaria:2018xxl, Beccaria:2020azj, Hellerman:2021yqz}. 

%In the particular case of ${\cal N}=4$ SU(N) SYM, one can also use holography to study these correlators. Are ALL extremal correlators non-renormalized for N=4?

In the current work, we will focus on the planar limit of some of these extremal correlators. For concreteness, we will present explicit results for single-trace operators of ${\cal N}=2$ SU(N) SYM with $\textnormal{N}_F=2 \textnormal{N}$ massless hypermultiplets in the fundamental representation, sometimes referred to as ${\cal N}=2$ SQCD. The techniques we will use, however, can be easily extended to any other Lagrangian ${\cal N}=2$ SCFT that admits a planar limit, and to correlation functions of multi-trace chiral operators. It was argued in \cite{Baggio:2014ioa} that extremal n-point functions are determined in terms of $2-$ and $3-$point functions, so we will restrict our attention to these. We obtain what we believe are the first known all-order analytic expressions for coefficients in the perturbative expansion of the planar limit of these 2- and 3- point functions.  {\it En route} to deriving these results, we deduce a combinatorial expression for the planar free energy of the relevant matrix model, and combinatorial expressions for the planar  2- and 3- point functions on $S^4$. In the rest of the introduction we briefly sketch a summary of these results and the methods used to derive them, and point out some possible extensions of the present work.

Four dimensional ${\cal N}=2$ SCFTs theories have various subsets of distinguished operators (see \cite{Dolan:2002zh} for a thorough discussion on ${\cal N}=2$ SCFTs short multiplets). In particular, chiral primary operators (CPOs) are defined as being annihilated by all  right chiral supercharges; similarly, anti-chiral operators are annihilated by all left chiral supercharges. CPOs have conformal dimension $\Delta$ fixed by their $U(1)_R$ R-charge , $\Delta =R/2$ and are $SU(2)_R$ singlets. Anti-chiral primary operators have $\Delta =-R/2$. We will consider CPOs that are Lorentz scalars, so they are characterized by their conformal dimension $\Delta$.  On $\bR^4$, correlation functions of CPOs and anti-CPOs can be non-zero only if the sum of their R-charges is 0. In particular, this implies that n-point functions of chiral primary operators (with no anti-chirals) are zero. The simplest non-trivial case are the extremal correlation functions, involving $n-1$ CPOs $O_i$ and a single anti-chiral operator $\bar O$
%\be
%\vev{O_{\Delta_1}(x_1) \dots O_{\Delta_{n-1}}(x_{n-1}) \bar O_{\bar \Delta}(y)} = \vev{O_{\Delta_1} \dots O_{\Delta_{n-1}} \bar O_{\bar \Delta}} (\tau,\bar \tau)
%\prod_{i=1}^{n-1} \frac{1}{|x_i-y|^{2\Delta_i}}
%\ee

\be
\vev{O_{\Delta_1}(x_1) \dots O_{\Delta_{n-1}}(x_{n-1}) \bar O_{\bar \Delta}(y)} = \frac{\vev{O_{\Delta_1} \dots O_{\Delta_{n-1}} \bar O_{\bar \Delta}} (\tau,\bar \tau)}{|x_1-y|^{2\Delta_1}\dots |x_{n-1}-y|^{2\Delta_{n-1}}}
\ee 
with $\Delta_1+\dots \Delta_{n-1}=\bar \Delta$. The position-independent coefficients $\vev{O_{\Delta_1} \dots O_{\Delta_{n-1}} \bar O_{\bar \Delta}} (\tau,\bar \tau)$ are non-holomorphic functions of the complexified coupling $\tau=\frac{2\theta}{\pi}+i\frac{4\pi}{g_{\text{YM}}^2}$ and their determination is the driving question for this work.

In this paper we restrict to Lagrangian SCFTs. The CPOs we will consider are single-trace operators involving the complex scalar $\phi$ of the ${\cal N}=2$ vector multiplet, $O_m  \, \propto \, \text{Tr } \phi^m$. $O_m$ has dimension $\Delta=m$. In the planar limit of theories with a single gauge coupling,  extremal 2- and normalized 3- point functions on $\bR^4$ are of the form
\be
\vev{O_k \widebar{O_k}} = k \left( \frac{\lambda}{16 \pi^2}\right)^k \left[1+\sum_{m=1}^\infty \sum_{n_1,\dots,n_m=2}^\infty a_k(n_1,\dots,n_m) \zeta_{2n_1-1}\dots \zeta_{2n_m-1} \left(\frac{\lambda}{16 \pi^2}\right)^{n_1+\dots+n_m} \right]
\label{generictwo}
\ee

\be
\frac{\vev{O_{k_1} O_{k_2} \bar{O}_{k_1+k_2}}_n}{\sqrt{k_1 \cdot k_2 \cdot (k_1+k_2)}} 
=\frac{1}{\textnormal{N}} \left[ 1+\sum_{m=1}^\infty \sum_{n_1,\dots,n_m=2}^\infty b_{k_1,k_2}(n_1,\dots,n_m) \zeta_{2n_1-1}\dots \zeta_{2n_m-1}  \left(\frac{\lambda}{16 \pi^2}\right)^{n_1+\dots+n_m} \right]
\label{genericthree}
\ee
with $\zeta_i$ values of the $\zeta$ function, $\lambda=g_{\text{YM}}^2 \textnormal{N}$ the 't Hooft coupling and $a_k(n_i)$ and $b_{k_1,k_2}(n_1,\dots,n_m)$ rational numbers. For ${\cal N}=2$ SQCD, we have computed $a_k(n)$ explicitly for $k=2,4,6$, and the expressions we find suggest the following conjecture
\be
\langle O_k \bar O_k\rangle \stackrel{?}{=} k\left(\frac{\lambda}{16\pi^2}\right)^k \left(1-2k \sum_{n=2}^\infty \frac{\zeta_{2n-1}}{n} \left(\frac{-\lambda}{16\pi^2}\right)^n {2n \choose n} \left[ (-1)^k {2n \choose n+k}+{2n \choose n+1}-n \right]+\dots \right)
\label{twopointintro}
\ee
where the dots stand for terms with products of two or more values of the $\zeta$ function. Similarly, we have computed $\vev{O_2O_2\bar O_4}_n$ and $\vev{O_2O_4\bar O_6}_n$ and the results obtained suggest the following conjecture for even $k_1,k_2$
\begin{equation}
\begin{split}
\frac{\langle O_{k_1} O_{k_2} \bar{O}_{k1+k2} \rangle_n}{\sqrt{k_1 \cdot k_2 \cdot (k_1+k_2)}} \stackrel{?}{=} \frac{1}{\textnormal{N}}&\Biggl[ 1-\sum_{n=2}^\infty \left(\frac{-\lambda}{16\pi^2}\right)^n \zeta_{2n-1} {2n \choose n} \\
&\left( {2n \choose n+k_1} + {2n \choose n + k_2}+{2n \choose n+k_1+k_2}+(n-1) (\mathcal{C}_n-2) \right)+\dots \Biggr]
\end{split}
\label{threepointintro}
\end{equation}
where again the dots stand for terms with products of two or more values of the $\zeta$ function and ${\cal C}_n$ are Catalan numbers. If we assign transcendality $n$ to $\zeta_n$ and in general $n_1+\dots+n_m$ to $\zeta_{n_1}\dots \zeta_{n_m}$, then at every order in the planar perturbative series, our conjectures refer to the term with maximal transcendality. The two analytic expressions we propose are strikingly simple, and certainly simpler than the intermediate results used to arrive at them. This suggests that there may be a more direct way to obtain them than the one pursued in this work. We will come back to this point at the end of the introduction.

The technical tool that we have used to derive (\ref{twopointintro}) and (\ref{threepointintro}) is supersymmetric localization \cite{Pestun:2007rz}. Supersymmetric localization has produced a plethora of exact results for supersymmetric quantum field theories in various dimensions (see \cite{Pestun:2016zxk} for a review) by reducing the evaluation of selected observables to matrix model computations. It is thus natural to try to apply it to the computation of chiral correlation functions. For CPOs with $\Delta=2$, it was argued in \cite{Gerchkovitz:2014gta} that this 2-point function can be obtained directly from the partition function of the CFT on $S^4$. For more general CPOs, the situation is more complicated: it was argued in \cite{Gerchkovitz:2016gxx} that correlation functions of CPOs on $S^4$ can be extracted from the $S^4$ partition function of a deformed theory. Furthermore, correlation functions on $S^4$ differ from those on $\bR^4$; to obtain the latter from the former, one needs to apply the Gram-Schmidt orthogonalization procedure  \cite{Gerchkovitz:2016gxx}.

The path described above has been followed already in a number of papers \cite{Baggio:2014ioa, Baggio:2015vxa, Baggio:2016skg, Rodriguez-Gomez:2016ijh, Rodriguez-Gomez:2016cem, Billo:2017glv, Beccaria:2020hgy,Galvagno:2020cgq, Beccaria:2021hvt}. The novel ingredient that we introduce in this work is an alternative way of evaluating the free energy and correlators of the relevant matrix models, which allows us to obtain all-order analytic expressions in the planar limit. Usually, Hermitian matrix model integrals are solved by reducing them to a Cartan subalgebra, which reduces the number of integrals, at the price of introducing a non-trivial Jacobian, the Vandermonde determinant. Instead, it is possible to tackle them in the original full Lie algebra formulation, an approach that in the context of supersymmetric localization has been pioneered in \cite{Billo:2017glv, Billo:2018oog, Billo:2019fbi, Fiol:2018yuc}. In this approach, the relevant matrix models for genuinely ${\cal N}=2$ SCFTs can be rewritten in terms of an action with infinitely many single and double trace terms \cite{Billo:2019fbi, Fiol:2020bhf, Fiol:2020ojn}. Furthermore, in the planar limit, it has been argued \cite{Fiol:2020bhf, Fiol:2020ojn} that the full perturbative series in $\lambda$ for various observables can be written in terms of a sum over tree graphs. In this work, when applying this strategy to the relevant matrix model, the main novelty compared to \cite{Fiol:2020bhf} is that now the single-trace terms in the matrix model action also contribute to the planar limit, complicating the analysis. Nevertheless, the resulting expressions for the planar free energies and correlation functions still involve sums over tree graphs.

This work leaves open a number of questions. First, it would be completely straightforward but rather tedious to extend the computations presented here to the terms involving a product of two values of $\zeta$ or higher in (\ref{generictwo}) and (\ref{genericthree}). In this work, we have focused on the terms with maximal transcendality; it might be possible to find analytic formulas for the coefficients of ofher terms with simple patterns, like those with just powers of $\zeta_3$, as in \cite{Galvagno:2021bbj}. It should also be possible to extend the analysis presented here for ${\cal N}=2$ SCQD to extremal correlators of other Lagrangian ${\cal N}=2$ SCFTs \cite{Fiol:2015mrp, Beccaria:2020hgy, Galvagno:2020cgq, Beccaria:2021hvt}. A very interesting problem would be to prove our conjectures (\ref{twopointintro}) and (\ref{threepointintro}). Conceivably, a proof might just extend our computations for arbitrary values of the conformal dimensions; after all, the relevant ingredients are the coefficients of the correlators on $S^4$, and the Gram-Schmidt relation to correlation functions on $\bR^4$, and both of these are known. A potentially more illuminating proof might bypass the relation to $S^4$ correlators, and work directly on $\bR^4$. Indeed, the factor $ \left(\frac{-\lambda}{16\pi^2}\right)^n {2n \choose n}  \frac{\zeta_{2n-1}}{n}$ that appears in (\ref{twopointintro}) and (\ref{threepointintro}) coincides with the values of a certain family of Feynman diagrams considered in closely related work \cite{Beccaria:2020hgy, Galvagno:2020cgq} (see also \cite{Mitev:2014yba}), so the form of (\ref{twopointintro}) and (\ref{threepointintro}) suggest that they can be proven by a combinatorial argument, counting the ways in which those particular Feynman diagrams enter the evaluation of $\vev{O_k \bar O_k}$ and $\vev{O_{k_1} O_{k_2} \bar O_{k_1+k_2}}$.

The structure of the paper is the following. In section 2 we consider a Hermitian matrix model with an action containing infinitely many single and double trace terms with arbitrary coefficients; we extend the analysis of \cite{Fiol:2020bhf}, and manage to write the planar free energy and the planar 2- and 3-point functions as sums over tree graphs. In section 3  we consider the evaluation of correlation functions of ${\cal N}=2$ SCFTs on $S^4$ through supersymmetric localization. We argue that the  relevant matrix model is a particular case of the one considered in section 2, thus obtaining expressions for the planar 2- and 3-point functions on $S^4$. Finally, in section 4 we apply the Gram-Schmidt procedure to the $S^4$ correlation functions found in the previous section, to obtain correlation functions on $\bR^4$. The manipulations become quite involved, thus preventing us from obtaining closed expressions for the full planar 2- and 3-point functions. Nevertheless, by focusing on the terms with a single value of $\zeta$, we compute them for operators of small conformal dimensions, and conjecture the formulas (\ref{twopointintro}) and (\ref{threepointintro}) for planar correlation functions of arbitrary single trace CPOs.

\section{Matrix Model with single and double traces}
One of the main technical tools that we will use in the following sections to compute extremal correlation functions of CPOs is supersymmetric localization. As we will argue, the resulting matrix models can be written in terms of an action involving infinitely many single and double trace term deformations, the latter having  very specific coefficients. In this section we study this type of matrix model with arbitrary coefficients, to highlight the generality of our arguments. 

Let's consider a Hermitian matrix model 
\begin{equation}
\mathcal{Z} = \int \textnormal{d}a \, e^{- \frac{1}{2g} \textnormal{Tr}(a^2)} e^{-S_{int}}
\end{equation}
where $a$ is a Hermitian $\textnormal{N}\times \textnormal{N}$ matrix, $\textnormal{d}a$ is the flat measure and $g$ is the matrix model coupling. The interacting part of the action consists of (possibly infinitely many) single and double trace terms,
\begin{equation}
S_{int}= \textnormal{N} \sum_{p\geq 3} c_p \textnormal{Tr } a^p \, + \, \sum_{mn} c_{mn} \textnormal{Tr } a^m \, \textnormal{Tr }a^n \, .
\label{def:st_dt_mm}
\end{equation}
with the coefficients $c_p,c_{mn}$  N-independent and otherwise arbitrary. Particular examples of this family of models have appeared in the study of two-dimensional quantum gravity \cite{Das:1989fq, Korchemsky:1992tt, AlvarezGaume:1992np, Klebanov:1994pv, Klebanov:1994kv}, and as reviewed in \cite{Grassi:2014vwa}, they have also appeared in many other contexts, from two-dimensional statistical mechanics, to three-dimensional gauge theories, or M-theory. Without the N factor in front of the single-trace terms, they are relevant \cite{Billo:2019fbi, Fiol:2020bhf} in the application of supersymmetric localization to four dimensional undeformed ${\cal N}=2$ super Yang-Mills theories. 

Our goals in this section are twofold: first, we will deduce the planar free energy for this family of models, as a function of the 't Hooft coupling and the coefficients $c_p,c_{ij}$. Then, in preparation for the next section, we will consider the coefficients $c_p$ as external sources; this will allow us to obtain the planar 2- and 3-point functions of single trace operators of the matrix model (\ref{def:st_dt_mm}) with just double-trace terms, by taking derivatives against the $c_p$ and then turning them off.

As shown in \cite{Grassi:2014vwa}, the planar free energy of these models can be deduced recursively, using the method of orthogonal polynomials. We will present an alternative expression for the planar free energy as a sum over tree graphs, generalizing \cite{Fiol:2020bhf}. More specifically, the matrix model considered in \cite{Fiol:2020bhf} was similar to (\ref{def:st_dt_mm}), but without the power of N in front of the single trace terms, rendering them irrelevant in the planar limit. On the other hand, the matrix model we will encounter in the next section is precisely of the form (\ref{def:st_dt_mm}). Nevertheless, we will show that the resulting planar free energy can still be written as a sum over tree graphs, albeit a more complicated one.

To study the planar limit of (\ref{def:st_dt_mm}), start by defining the matrix model 't Hooft coupling by $\tlambda=g \textnormal{N}$\footnote{This matrix model 't Hooft coupling $\tlambda$ differs by a constant from the Yang-Mills 't Hooft coupling $\lambda = 16\pi^2 \tlambda$, to be introduced in the next section.}. In the large N limit, the free energy of the matrix model admits an expansion of the form
\begin{equation}
\mathcal{F}(\tlambda, N) = - \log \mathcal{Z} = - \sum_{m=1}^\infty \frac{(-1)^{m+1}}{m}\left( \sum_{k=1}^\infty \frac{1}{k!}\vev{(-S_{int})^k} \right)^m = F_0(\tlambda) N^2 + \cdots \, ,
\label{def:fenergy}
\end{equation}
where the $\textnormal{N}^2$ contribution is given by the planar free energy $F_0(\tlambda)$. At a given order we will have all the possible factorizations of a general correlator of the form $\langle S_{int}^m \rangle$, thus the terms contributing to the planar free energy will be those that scale as N$^2$ and survive all the cancellations arising from the logarithm. We wish then to characterize this set of terms. In particular, let's consider a term with $m-k$ single traces and $k$ double trace; its contribution is of the form
\begin{equation}
\textnormal{N}^{m-k} \langle \textnormal{Tr } a^{p_1} \dots \textnormal{Tr } a^{p_{m-k}} \textnormal{Tr } a^{m_1} \textnormal{Tr } a^{n_1} \dots \textnormal{Tr } a^{m_k} \textnormal{Tr } a^{n_k} \rangle ,
\end{equation}
Since the planar free energy scales like $\textnormal{N}^2$, we want to extract the part of the correlator that scales like $\textnormal{N}^{2-m+k}$. Consider a contribution given by the product of $s$ connected correlators of sizes $r_1,\dots, r_s$. This product scales like $\textnormal{N}^{2s-(r_1+\dots+r_s)}$, and since $r_1+\dots +r_s=m+k$, we find that $s=k+1$.

We have learned that when there are $k$ double traces in $\langle S^m\rangle$, the terms that scale as $\textnormal{N}^2$ are products of $k+1$ connected correlators. As in  \cite{Fiol:2020bhf}, we can associate a graph to this product of connected correlators, with one vertex per correlator and one edge per double trace. Not all these terms contribute to the free energy, they must survive the logarithm. The argument from \cite{Fiol:2020bhf} still goes through, and the terms that survive are those whose graph is a tree, see  \cite{Fiol:2020bhf} for the details of the argument.

However, there are various differences with respect to the case of a potential with just double traces. Now at order $\langle S^m\rangle$ we must consider trees with $0\leq k\leq m$ edges. The case $k=0$ corresponds to the connected correlator of just single traces; the term $k=m$ corresponds to the case with just double traces. So, for fixed $k$, we must sum over all the ways to distribute the $m-k$ single traces in the $k+1$ correlators. More explicitly, the planar Free Energy for the family of theories such as (\ref{def:st_dt_mm}) is given by\footnote{Note that we are not including the free energy of the Gaussian model in this expression.}

\be
\begin{split}
{\cal F}=\sum_{m=1}^\infty \frac{(-1)^m}{m!} \sum_{k=0}^m {m \choose k} &\sum_{p_1,\dots,p_{m-k}} c_{p_1}\dots c_{p_{m-k}} \sum_{\substack{i_1,\dots,i_k \\ j_1,\dots, j_k}} c_{i_1j_1}\dots c_{i_kj_k} \\
&\sum_{\substack{\text{directed trees with} \\ \text{k labeled edges}}} \sum_{\substack{\text{single trace} \\ \text{insertions}}}  \prod_{i=1}^{k+1} V_i
\label{res:free_e}
\end{split}
\ee
where $V_i$ is the planar connected correlator on the $i$-th vertex on the tree, that contains the following operators: tr $a^{i_s}$ if the directed edge labelled $s$ leaves that vertex; tr $a^{j_s}$ if the directed edge labelled $s$ arrives at that vertex; any single trace operators inserted on that vertex. 

It is worth comparing this result with the one obtained in  \cite{Fiol:2020bhf}, valid for potentials with only double trace contributions in the large N limit. First, as already mentioned, now the sum at order $m$ involves trees with $k\leq m$ edges. Second, in the case of just double-trace terms in the action, it is easy to argue \cite{Fiol:2020bhf} that double traces of odd powers don't contribute to the planar limit. The argument was based on the fact that a planar connected correlator must involve an even number of odd powers. However, the argument doesn't apply now, because the single traces in (\ref{def:st_dt_mm}) can also have odd powers. So  (\ref{res:free_e}) includes also contributions coming from double traces of odd powers.

We have succeeded in writing the planar free energy as a sum over products of planar connected correlators of the free Gaussian model. To proceed, we need the explicit form of these planar connected correlators. They are known in some cases, but not all. For an arbitrary n-point function of even-power operators
 \cite{tutte, Gopakumar:2012ny} 
\begin{equation}
	\langle \textnormal{Tr}a^{2 k_1} \cdots \textnormal{Tr}a^{2 k_n} \rangle_c = \tlambda^d \frac{(d-1)!}{(d-n+2)!} \prod_{i=1}^n \frac{(2k_i)!}{k_i! (k_i-1)!} \textnormal{N}^{2-n}
	\label{def:gop_even}
\end{equation}
where $d = \sum k_i$. Let us introduce some notation for the numerical coefficients 
\begin{equation}
{\cal V } (k_1,\cdots, k_n) =  \frac{(d-1)!}{(d-n+2)!} \prod_{i=1}^n \frac{(2k_i)!}{k_i! (k_i-1)!} \, .
\label{def:nu_coef}
\end{equation}

The planar 2-point function of odd operators is
\begin{equation}
\langle \textnormal{Tr} a^{2k_1+1} \textnormal{Tr} a^{2k_2+1} \rangle_c = \frac{\tlambda^{k_1+k_2+1}}{k_1+k_2+1}\frac{(2k_1+1)!}{(k_1!)^2}\frac{(2k_2+1)!}{(k_2!)^2}
\label{def:gop_odd}
\end{equation}
A general formula is also known for the case of correlators involving and arbitrary number of even operators with two odd insertions \cite{tutte, Gopakumar:2012ny}. Finding the generalization to more than two odd insertions is an interesting open question. 

As a check of (\ref{res:free_e}), consider the expansion up to $m=2$ in the case of just even traces

\be
\begin{split}
{\cal F}=&-\sum_p c_{2p}  \frac{(2p)!}{(p+1)! p!}\tilde{\lambda}^p -\sum_{ij}c_{2i\, 2j}  \frac{(2i)!(2j)!}{(i+1)!i! (j+1)! j!}\tilde{\lambda}^{i+j}+\frac{1}{2}\sum_{p,q}  \frac{c_{2p}\, c_{2q}}{p+q}\frac{(2p)!(2q)!}{(p-1)! p! (q-1)! q!}\tilde{\lambda}^{p+q}\\
&+2\sum_p \sum_{ij}  \frac{c_{2p}\, c_{2i\, 2j}}{p+i} \frac{(2p)!}{(p-1)! p!} \frac{(2i)!}{(i-1)! i!}\frac{(2j)!}{(j+1)! j!} \tilde{\lambda}^{p+i+j} \\
&+2\sum_{i,j,k,l}\frac{c_{2i,2j}\, c_{2k,2l}}{j+k}\frac{(2i)!}{(i+1)! i}\frac{(2j)}{(j-1)!j!}\frac{(2k)!}{(k-1)! k!}\frac{(2l)!}{(l+1)! l!} \tilde{\lambda}^{i+j+k+l}+\dots
\end{split}
\ee
this model is now the one in appendix B of \cite{Grassi:2014vwa} and the expression above reproduces all the relevant terms in \cite{Grassi:2014vwa}.

\subsection{$2-$ and $3-$point functions}
In preparation for the next section, we now compute the planar 2- and 3- point functions of the matrix model with interaction terms (\ref{def:st_dt_mm}).  Note that the expression (\ref{res:free_e}) contains a sum over directed trees with $k$ labeled edges and by taking a derivative with respect to any coupling we are selecting from the sum the trees that contain, in one of the vertices, an insertion of the operator associated to the aforementioned coupling. This distinguishes one of the vertices from the rest, turning the tree into a rooted tree, where the root vertex indicates  the correlator that contains the selected operator. In the case of higher point functions we will have as many roots as operators we wish to consider, while bearing in mind that we can have multiple roots in the same vertex of the tree. \footnote{This is similar to a coloring of a given tree, but in that case it is not possible to paint the same vertex with multiple different colors.}

Let us be more explicit for the correlation functions that we are interested in.  For $2-$point functions (\ref{res:free_e}) reduces to
%\begin{equation}
%\langle \textnormal{Tr}a^{p} \textnormal{Tr}a^{q} \rangle = 2 \sum_{m=2}^\infty \frac{(-1)^m}{m!}{m \choose m-2} \sum_{\substack{i_1,\dots,i_{m-2} \\ j_1,\dots, j_{m-2}}} c_{i_1 j_1}\cdots c_{i_{m-2}j_{m-2}} \sum_{\substack{\text{double rooted} \\ \text{directed trees} \\ \text{with m-2 labeled edges}}} \prod_{i=1}^{m-1} {\cal V}_i \, ,
%\label{res:2pt_stdt_mm}
%\end{equation}

\begin{equation}
\langle \textnormal{Tr}a^{p} \textnormal{Tr}a^{q} \rangle =\sum_{m=0}^\infty \frac{(-1)^m}{m!}\sum_{\substack{i_1,\dots,i_m \\ j_1,\dots, j_{m}}} c_{i_1 j_1}\cdots c_{i_{m}j_{m}} \sum_{\substack{\text{double rooted} \\ \text{directed trees} \\ \text{with m labeled edges}}} \prod_{i=1}^{m+1} V_i \, ,
\label{res:2pt_stdt_mm}
\end{equation}
with the understanding that the tree with $m=0$ edges is just a single vertex, corresponding to the connected Gaussian two-point function, $\langle \textnormal{Tr}a^{p} \textnormal{Tr}a^{q} \rangle_c$. The two distinguished vertices - roots - of the tree correspond to the insertions of Tr $a^p$ and Tr $a^q$ (they can be inserted in the same vertex). The derivation of this formula guarantees its validity for $p>2, q>2$, but it is possible to check that it is also valid for $p=2$ and/or $q=2$ by using this relation
\begin{equation}
\langle  \textnormal{Tr}a^{2}  \textnormal{Tr}a^{2k_2}\dots  \textnormal{Tr}a^{2k_n}\rangle_c = 
\frac{2}{\textnormal{N}}\tlambda^2 \partial_{\tlambda} 
\langle  \textnormal{Tr}a^{2k_2}\dots  \textnormal{Tr}a^{2k_n}\rangle_c 
\end{equation}
which follows from (\ref{def:gop_even}).
To illustrate (\ref{res:2pt_stdt_mm}), let's compute the first terms. We assume that $c_{pq}=c_{qp}$. In the even-even case, $\vev{ \textnormal{Tr}a^{2m} \textnormal{Tr}a^{2n}}$ the first non-trivial contribution comes from two types of products of planar connected Gaussian correlators: $\vev{ \textnormal{Tr}a^{2m} \textnormal{Tr}a^{2n} \textnormal{Tr}a^{2p}}_c \vev{\textnormal{Tr}a^{2q}}$, and $ \vev{ \textnormal{Tr}a^{2m} \textnormal{Tr}a^{2p}}_c \vev{ \textnormal{Tr}a^{2q} \textnormal{Tr}a^{2n}}_c$. Both types of products correspond to a tree with a single edge and two vertices; in the first case,  the two single trace operators are both inserted in the same vertex, and in the second case, each single trace operator in inserted in a different vertex. All in all, 
\begin{equation}
\begin{split}
&\vev{ \textnormal{Tr}a^{2m} \textnormal{Tr}a^{2n}}=\frac{1}{m+n} \frac{(2m)!}{(m-1)! m!} \frac{(2n)!}{(n-1)! n!} \tlambda^{m+n} \\
&- 2 \sum_{p,q} c_{2p,2q} \frac{\tlambda^{p+q+n+m} (2p)!(2q)!(2m)!(2n)!}{p!(p-1)!q!(q-1)!m!(m-1)!n!(n-1)!} \left(\frac{1}{(p+1)p}+\frac{1}{(p+n)(q+m)}\right) + \cdots
\end{split}
\label{res:ee_s4}
\end{equation}
The odd-odd two-point function works similarly, with the difference that now when both single trace insertions are in different correlators, the double trace must be odd-odd, and if they are in the same correlator, the double-trace must be even-even,
\begin{equation}
\begin{split}
\vev{\textnormal{Tr}a^{2m+1}\textnormal{Tr}a^{2n+1}} &= \frac{\tlambda^{m+n+1}}{m+n+1}\frac{(2m+1)!(2n+1)!}{(
m!)^2(n!)^2}\\
&-2  \tlambda^{m+n+1}\frac{(2m+1)!(2n+1)!}{n!^2 m!^2} \sum_{ij}\Biggl( c_{2i+1,2j+1}\frac{\tlambda^{i+j+1}(2i+1)!(2j+1)!}{(m+i+1)(i!)^2(n+j+1)(j!)^2} \\
&+ c_{2i,2j} \frac{\tlambda^{i+j}(2i)!(2j)!}{(i!)(i-1)!(j+1)!j!} \Biggr) + \cdots 
\end{split}
\label{res:oo_s4}
\end{equation}
In the case of $3-$point functions we have
%\begin{equation}
%\langle \textnormal{Tr}a^{p} \textnormal{Tr}a^{q} \textnormal{Tr}a^{l}\rangle = 6 \sum_{m=3}^\infty \frac{(-1)^m}{m!}{m \choose m-3} \sum_{\substack{i_1,\dots,i_{m-3} \\ j_1,\dots, j_{m-3}}} c_{i_1 j_1}\cdots c_{i_{m-3}j_{m-3}} \sum_{\substack{\text{triple rooted} \\ \text{directed trees} \\ \text{with m-3 labeled edges}}} \prod_{i=1}^{m-2} {\cal V}_i
%\label{res:3pt_stdt_mm}
%\end{equation}

\begin{equation}
\langle \textnormal{Tr}a^{p} \textnormal{Tr}a^{q} \textnormal{Tr}a^{l}\rangle = \textnormal{N}^{-1}\sum_{m=0}^\infty \frac{(-1)^m}{m!}  \sum_{\substack{i_1,\dots,i_{m} \\ j_1,\dots, j_{m}}} c_{i_1 j_1}\cdots c_{i_{m}j_{m}} \sum_{\substack{\text{triple rooted} \\ \text{directed trees} \\ \text{with m labeled edges}}} \prod_{i=1}^{m+1} V_i
\label{res:3pt_stdt_mm}
\end{equation}
This formula applies to the two non-trivial cases: three even powers, and two odd powers and an even one. Let's illustrate it with the first case,
\begin{equation}
\begin{split}
& \langle \textnormal{Tr}a^{2p} \textnormal{Tr}a^{2q} \textnormal{Tr}a^{2l}\rangle = \textnormal{N}^{-1}
\frac{(2p)!(2q)!(2l)!}{p! (p-1)! q! (q-1)! l! (l-1)!} \tlambda^{p+q+l} \\
& \left[1-2 \sum_{ij}c_{2i,2j} \tlambda^{i+j} \frac{(2i)!(2j)!}{i!(i-1)! j! (j-1)!} \left(\frac{p+q+l+i-1}{(j+1)j}+\frac{1}{p+j}+\frac{1}{q+j}+\frac{1}{l+j} \right)+\dots\right]
\end{split}
\end{equation}

In the next section, we will evaluate these generic expressions for the specific matrix model of ${\cal N}=2$ SQCD. As we will see, they reproduce and generalize known results \cite{Galvagno:2020cgq}, thus providing a non-trivial check of their validity.

\section{Chiral Correlators on $S^4$}

In this section we derive planar 2- and 3- point functions of single-trace chiral primary operators of ${\cal N}=2$ SQCD on $S^4$, using the results derived in the previous section.

Let us first quickly recall some basic facts about 4d ${\cal N}=2$ SCFT theories and their chiral primary operators \cite{Dolan:2002zh}. The generators of the superconformal algebra are given by the bosonic generators $P_\mu,K_\mu,M_{\mu \nu}, D$, the supercharges $Q^a_{\alpha}, \bar{Q}^a_{\dot{\alpha}}$, its superconformal partners $S^a_{\alpha}, \bar{S}^a_{\dot{\alpha}}$ and the generators of the $SU(2)\times U(1)$ R-symmetry. Highest weight representations are labelled by the quantum numbers $(\Delta; j_l,j_r;s;R)$ of the highest weight state under dilatations, the Lorentz group and the $SU2)\times U(1)$ R-symmetry group. These highest weight states are created by superconfornal primary operators, annihilated by all $S^a_{\alpha}, \bar{S}^a_{\dot{\alpha}}$.

Among all of the superconformal primaries, there exists an interesting class given by the ones that are chiral, defined as $[\bar{Q}^a_{\dot{\alpha}},O]=0$. CPOs have $j_r=s=0$ and $\Delta=R/2$. For Lagrangian SCFTs, one can further argue that $j_l=0$, so they are Lorentz scalars \cite{Buican:2014qla}.  Anti-chiral primary operators $\bar O$ are similarly defined, and satisfy $\Delta=-R/2$.
We will denote chiral operators on $S^4$ by $\Omega$, reserving $O$ for chiral operators on $\bR^4$.

\subsection{${\cal N}=2$ SCFTs on $S^4$}

It is possible to place any ${\cal N}=2$ SCFT on $S^4$. Supersymmetric regularization of the resulting divergences implies that the theory preserves a subalgebra $osp(2|4)$ of the flat space supersymmetry algebra \cite{Gomis:2014woa}. In particular, the flat space $U(1)_R$ symmetry is broken on $S^4$  \cite{Gomis:2014woa}. As a consequence, there is no $U(1)_R$ selection rule for correlation functions on $S^4$: one-point functions are not vanishing, and similarly, two-point functions of operators of different dimension can also be non-zero. In \cite{Gerchkovitz:2014gta} it was shown that the partition function on $S^4$ can be identified with the K\"ahler potential for the Zamolodchikov metric of the conformal manifold of the theory. Thus, the two-point function of CPOs with $\Delta=2$ can be obtained by taking derivatives of this partition function on $S^4$.

Supersymmetric localization allows to evaluate efficiently this partition function, and consequently this very particular two-point function of CPOs, by reducing it to a matrix model integral \cite{Pestun:2007rz},
\begin{equation}
		Z_{S^4} (\tau_{\text{YM}})= \int da \, e^{-\frac{8\pi^2}{g_{\text{YM}}^2} \textnormal{Tr}(a^2)} \mathcal{Z}_{1-loop} (a) \, |\mathcal{Z}_{inst}(a,\tau)|^2
\label{zs4intro}		
\end{equation} 
where $\mathcal{Z}_{1-loop}$ a factor arising from a $1-$loop computation and $\mathcal{Z}_{inst}$ is the instanton contribution, that it is usually assumed to be negligible in the large N limit.

These results were extended in \cite{Gerchkovitz:2016gxx}, where a method  to exactly compute correlation functions of chiral primary operators on $S^4$ was developed. The starting point is to consider a deformation of the theory on $S^4$ that involves new couplings, one per generator of the chiral ring of the theory. This deformed SCFT still preserves $osp(2|4)$. It was argued in \cite{Gerchkovitz:2016gxx} that extremal correlators on $S^4$ of the undeformed theory can be obtained by taking derivatives of the partition function of the deformed theory. 

Again, supersymmetric localization allows to efficiently evaluated this new partition function, and thus arbitrary extremal correlators. Indeed, it was proven in \cite{Gerchkovitz:2016gxx} that the deformed partition function can be obtained from a matrix model integral of the form
\begin{equation}
Z_{S^4} = \int da \,  \lvert e^{i \sum_{n=1}^m \pi^{n/2} \tau_n \textnormal{Tr}(a^n)}\lvert ^2 \mathcal{Z}_{1-loop}(a) \, |\mathcal{Z}_{inst} (a,\tau, \tau_n)|^2
\end{equation}
with $\tau_n$ a holomorphic coupling and $\tau_2 = \tau_{\text{YM}} = \frac{\theta}{2\pi}+ \frac{4\pi i}{g^2}$. Note that the 1-loop partition function does not depend on the new couplings $\tau_n$, but the instanton partition function does. The key point is that correlation functions of chiral operators on $S^4$ are given by correlation functions of this matrix model,
\begin{equation}
\vev{\Omega_p \Omega_q}_{S^4}= \vev{ \textnormal{Tr}a^{p}  \, \textnormal{Tr}a^{q}}_{\text{MM}}
\end{equation}
and similarly for higher n-point functions. This fixes the normalization of the CPOs.

In this work, we focus on the large N limit of these correlators, and that implies a number of simplifications: first, we can restrict the terms we add to the action to single-trace CPOs; second, we will neglect 
the instanton contribution, setting $\mathcal{Z}_{inst} (a,\tau, \tau_n)=1$. We thus rewrite the deformation as
	\begin{equation}
		S = - i \sum_{n=2}^m \pi^{\frac{n}{2}} (\tau_n-\bar{\tau}_n) \textnormal{Tr}a^n = \frac{8\pi^2}{g^2} \textnormal{Tr}a^2 - i \sum_{n=3}^m \pi^{\frac{n}{2}}(\tau_n-\bar{\tau}_n) \textnormal{Tr}a^n \, ,
		\label{eq:stdeform}
	\end{equation}
where now we can identify $g = \frac{g_{\text{YM}}^2}{16\pi^2}$ and we recognize the single trace deformation to be the one in (\ref{def:st_dt_mm}). Following \cite{Billo:2017glv, Billo:2018oog,Billo:2019fbi, Fiol:2020bhf} it is possible to rewrite $S$ as a sum of single and double trace terms. For ${\cal N}=2$ SQCD
\begin{equation} 
S_{int} =\sum_{n=2}^\infty \frac{\zeta(2n-1)(-1)^n}{n} \left[  \sum_{k=1}^{n-1}{2n \choose 2k} \textnormal{Tr}a^{2(n-k)} \textnormal{Tr}a^{2k} 
 - \sum_{k=1}^{n-2}{2n \choose 2k+1} \textnormal{Tr}a^{2(n-k)-1} \textnormal{Tr}a^{2k+1} \right],
\label{eq:n2_action}
\end{equation} 
To sum up, the deformation of the ${\cal N}=2$ SCFT, together with the rewriting of the 1-loop determinant as an effective action, show that the relevant matrix model is of the type (\ref{def:st_dt_mm}) analyzed in the previous section.

%{\it Normalization of the operators. Ordinarily, in CFT the normalization of operators is fixed by requiring that operators are orthonormal with regards to their two-point functions. However, in the literature of extremal correlation functions is more common to fix the normalization by requiring that the OPE product is of the form [THIS FIXES THE NORMALIZATION OF MULTI-TRACE OPERATORS, BUT NOT OF SINGLE TRACE OPERATORS]
%\be
%O_m(x) O_n(0) = O_{m+n}(0)+\dots
%\ee
%It is important to keep in mind that these single-trace operators are not the only CPOs of a given dimension $\Delta$. This leads to operator mixing. }

\subsection{Chiral correlators in $\mathcal{N}=4$}

As a warm-up, let's first recover the planar chiral 2- and 3- point functions of ${\cal N}=4$ $SU(\textnormal{N})$ SYM on $S^4$ with our techniques. In this case, supersymmetric localization reduces to the Gaussian matrix model, since the one-loop and the instanton contributions are trivial,  $\mathcal{Z}_{1-loop}=1$, $\mathcal{Z}_{inst}=1$. Thus, the planar 2- and 3-point functions on $S^4$ are just particular cases of (\ref{def:gop_even}) and (\ref{def:gop_odd}). Recalling the relation $16\pi^2 \tlambda  =\lambda$ between the matrix model and the Yang-Mills 't Hooft couplings, we have

\begin{equation}
\begin{split}
\vev{\Omega_{2n} \Omega_{2m}} &= \left( \frac{\lambda}{16\pi^2} \right)^{n+m}  \frac{1}{n+m}\frac{(2m)!}{m!(m-1)!}\frac{(2n)!}{n!(n-1)!} \\
\vev{\Omega_{2n+1} \Omega_{2m+1}} &= \left( \frac{\lambda}{16 \pi^2} \right)^{m+n+1} \frac{1}{m+n+1}\frac{(2m+1)!}{(m!)^2}\frac{(2n+1)!}{(n!)^2}
\end{split}
\end{equation}

\begin{equation}
\vev{\Omega_{2m}\Omega_{2n}\Omega_{2p}} = \left( \frac{\lambda}{16\pi^2} \right)^{m+n+p} \frac{(2m)!}{m!(m-1)!}\frac{(2n)!}{n!(n-1)!} \frac{(2p)!}{p! (p-1)!} \textnormal{N}^{-1}
\end{equation}
which agrees with the results obtained in  \cite{Rodriguez-Gomez:2016ijh}.

\subsection{Chiral operators in truly $\mathcal{N}=2$ theories}
Turning now our attention to truly $\mathcal{N}=2$ theories, we can identify the coefficients $c_{ij}$ in (\ref{def:st_dt_mm}) with the ones appearing in the effective action (\ref{eq:n2_action}) as 
\begin{equation}
c_{pq}= {2p+2q \choose 2p} \frac{\zeta(2p+2q-1) (-1)^{p+q}}{p+q}
\label{def:coef_n2}
\end{equation}
For Lagrangian $\mathcal{N}=2$ SCFTs theories, the planar free energy (\ref{res:free_e}) was explicitly computed in \cite{Fiol:2020bhf}, and for ${\cal N}=2$ SQCD it was found to be given by 
\begin{multline}
F_0(\lambda)=\frac{1}{2}\log\lambda+\sum_{n=2}^\infty \left(-\frac{\lambda}{16\pi^2}\right)^n \sum_{\substack{\text{compositions of n}\\ \text{not containing 1}}} (-2)^m 
\frac{\zeta (2n_1-1)\dots \zeta(2n_m-1)}{n_1\dots n_m} \\
\sum _{k_1=1}^{n_1-1} {2n_1 \choose 2k_1} \dots
\sum _{k_m=1}^{n_m-1} {2n_m \choose 2k_m}  
\sum_{\substack{\text{unlabeled trees} \\ \text{with m edges}}} \frac{1}{|\text{Aut(T)}|}  {\cal V}_1 \dots {\cal V}_{m+1}
\label{finalfree}
\end{multline}
Let us first note that we can obtain $\vev{\Omega_2 \Omega_2}$ by taking two derivatives of the free energy with respect to the exactly marginal coupling $g_{\text{YM}}$ of the theory. We obtain
\begin{multline}
\vev{\Omega_2 \Omega_2}=\frac{2\lambda^2}{(4\pi)^4}+\frac{4\lambda^2}{(4\pi)^4}\sum_{n=2}^\infty n(n+1) \left(-\frac{\lambda}{16\pi^2}\right)^n \sum_{\substack{\text{compositions of n}\\ \text{not containing 1}}} (-2)^m 
\frac{\zeta (2n_1-1)\dots \zeta(2n_m-1)}{n_1\dots n_m} \\
\sum _{k_1=1}^{n_1-1} {2n_1 \choose 2k_1} \dots
\sum _{k_m=1}^{n_m-1} {2n_m \choose 2k_m}  
\sum_{\substack{\text{unlabeled trees} \\ \text{with m edges}}} \frac{1}{|\text{Aut(T)}|}  {\cal V}_1 \dots {\cal V}_{m+1}
\label{res:final_22}
\end{multline}
where by expanding we see that the first terms match with eq. (4.13) of \cite{Galvagno:2020cgq}.

For the general planar 2- and 3-point functions on $S^4$ we can now use the results derived last section from the matrix model, eqs. (\ref{res:2pt_stdt_mm}), (\ref{res:3pt_stdt_mm})

\begin{equation}
\langle  \Omega_p \Omega_ q \rangle =\sum_{m=0}^\infty \frac{(-1)^m}{m!}\sum_{\substack{i_1,\dots,i_m \\ j_1,\dots, j_{m}}} c_{i_1 j_1}\cdots c_{i_{m}j_{m}} \sum_{\substack{\text{double rooted} \\ \text{directed trees} \\ \text{with m labeled edges}}} \prod_{i=1}^{m+1} V_i \, ,
\label{res:2pt_stdt_s4}
\end{equation}

\begin{equation}
\langle \Omega_p \Omega_q \Omega_l \rangle = \textnormal{N}^{-1}\sum_{m=0}^\infty \frac{(-1)^m}{m!}  \sum_{\substack{i_1,\dots,i_{m} \\ j_1,\dots, j_{m}}} c_{i_1 j_1}\cdots c_{i_{m}j_{m}} \sum_{\substack{\text{triple rooted} \\ \text{directed trees} \\ \text{with m labeled edges}}} \prod_{i=1}^{m+1} V_i
\label{res:3pt_stdt_s4}
\end{equation}

%The second case of interest are two-point functions of the form $\vev{\Omega_2 \Omega_{2p}}$, with $p > 1$. Now we need to consider the modified matrix model, but the only terms in the planar free energy (\ref{res:free_e}) that contribute to this $2-$point function are linear in the $\tau_p$ coupling. They correspond to trees with $\Omega_{2p}$ inserted in one of the vertices. This marks the vertex it is inserted in, so these are rooted trees, 
%\begin{equation}
%\begin{split}
%\vev{\Omega_2 \Omega_{2p}}= \frac{-2\lambda}{(4\pi)^2} \sum_{m=1}^\infty \frac{(-1)^m}{m!} {m \choose m-1} &\sum_{\substack{i_1,\dots,i_{m-1} \\ j_1,\dots, j_{m-1}}} c_{i_1 j_1}\cdots c_{i_{m-1}j_{m-1}}(i_1+j_1+\cdots +i_{m-1}+j_{m-1}+p) \\ 
% &\sum_{\substack{\text{rooted directed trees} \\ \text{with m-1 labeled edges}}} \prod_{i=1}^m {\cal V}_i
%\end{split}
%\end{equation}
%with the understanding that $\Omega_{2p}$ is inserted in the root vertex. \textcolor{red}{Is it possible to simplify the sum to undirected unlabeled trees?}.

While in most of this work we explicitly display terms with a single value of the $\zeta$ function, our formulas capture also all terms with products of two or more values of $\zeta$. To illustrate this point (see the Appendix for further examples), let's compute the $\zeta_3^2$ term for the 2-point function of two even single trace CPOs, 
\begin{equation}
\begin{split}
\vev{ \Omega_{2m} \Omega_{2n}} = 
&\frac{1}{m+n} \frac{(2m)!}{(m-1)! m!} \frac{(2n)!}{(n-1)! n!} \left(\frac{\lambda}{16 \pi^2}\right)^{m+n} \\
&-12 \zeta_3 \left(\frac{\lambda}{16 \pi^2}\right)^{m+n+2} \frac{(mn+m+n+3) (2m)! (2n)!}{(m+1)! (m-1)! (n+1)! (n-1)!} \\
&+ 72 \zeta_3^2 \left(\frac{\lambda}{16 \pi^2}\right)^{m+n+4}\frac{(3+m+n)(5+m+n+m n) (2m)! (2n)!}{(m+1)! (m-1)! (n+1)! (n-1)!}
\end{split}
\end{equation}
which agrees with (4.33), (4.34), (4.35) of \cite{Rodriguez-Gomez:2016ijh}.

\section{Chiral correlators on $\mathbb{R}^4$}
In the previous section, we have provided combinatorial expressions for the full planar perturbative series of 2- and 3- point functions of ${\cal N}=2$ superconformal theories on $S^4$.  As discussed above, it is not straightforward to read off the chiral correlators on $\mathbb{R}^4$ directly from the previous results. In order to do so we need to disentangle the mixing induced by the conformal anomaly through a Gram-Schmidt orthogonalization procedure \cite{Gerchkovitz:2016gxx}. In general, a given operator of dimension $\Delta_n$  will mix with all the operators with $\Delta_m$ such that $m < n$ differs from $n$ by an even integer. To find the relation between $\bR^4$ and $S^4$ operators, first introduce the matrix of two point functions on $S^4$ defined by

\begin{equation}
C_{n,m} = \vev{\Omega_n \Omega_m},
\label{def:2pt_matrix}
\end{equation}
then, the $\bR^4$ operator $O_n$ is given by

\begin{equation}
O_n(a) = \Omega_n(a) - \sum_{p,q} C_{n,p} \left( C^{-1}_{(n)} \right)^{p,q} \Omega_q(a) \, ,
\label{def:GS}
\end{equation}

While in the previous section we provided a combinatorial expression for planar chiral correlators on $S^4$, for the analogous correlators on $\bR^4$ obtained through the Gram-Schmidt procedure, a combinatorial description is no longer apparent. As discussed before, correlators of $\Delta= 2$ operators can be extracted directly from the partition function of $\mathcal{N}=2$ theories \cite{Gerchkovitz:2014gta}  which in turn admits an exact combinatorial expression \cite{Fiol:2020bhf} so let's start discussing 2-point functions on $\bR^4$. 

On $\bR^4$ the $U(1)_R$ selection rule implies that the only non-zero two-point functions are $\vev{O_k \bar O_k}$. Since $\vev{O_2 \bar O_2}$ is the same as on $S^4$, in this case we do have a full planar perturbative expression
\begin{multline}
\vev{O_2 \bar O_2}= \frac{4\lambda^2}{(4\pi)^4}\sum_{n=2}^\infty n(n+1) \left(-\frac{\lambda}{16\pi^2}\right)^n \sum_{\substack{\text{compositions of n}\\ \text{not containing 1}}} (-2)^m 
\frac{\zeta (2n_1-1)\dots \zeta(2n_m-1)}{n_1\dots n_m} \\
\sum _{k_1=1}^{n_1-1} {2n_1 \choose 2k_1} \dots
\sum _{k_m=1}^{n_m-1} {2n_m \choose 2k_m}  
\sum_{\substack{\text{unlabeled trees} \\ \text{with m edges}}} \frac{1}{|\text{Aut(T)}|}  {\cal V}_1 \dots {\cal V}_{m+1}
\end{multline}
In the Appendix we present the first few orders, that match with the results eq. (4.13) of \cite{Galvagno:2020cgq}. For future reference, we note that all the terms in $\langle O_2 \bar O_2\rangle$ with a single value of the $\zeta$ function can be rewritten as follows
\be 
\langle O_2 \bar O_2\rangle = 2\left(\frac{\lambda}{16\pi^2}\right)^2 \left(1-4 \sum_{n=2}^\infty \frac{\zeta_{2n-1}}{n} \left(\frac{-\lambda}{16\pi^2}\right)^n {2n \choose n} \left[ {2n \choose n+2}+{2n \choose n+1}-n \right]+\dots \right)
\label{o2o2r4}
\ee

% In a similar fashion the $3-$point function of chiral operators of dimension $2$ does not mix and thus we obtain
% \begin{equation}
% \begin{split}
% \langle O_2 O_2 O_2 \rangle = \textcolor{red}{\frac{-8\lambda^3}{(4\pi)^6}}\sum_{n=2}^\infty n(n+1)(n+2) \left(-\frac{\lambda}{16\pi^2}\right)^n \sum_{\substack{\text{compositions of n}\\ \text{not containing 1}}} (-2\beta_G)^m 
% \frac{\zeta (2n_1-1)\dots \zeta(2n_m-1)}{n_1\dots n_m} \\
% \sum _{k_1=1}^{n_1-1} {2n_1 \choose 2k_1} \dots
% \sum _{k_m=1}^{n_m-1} {2n_m \choose 2k_m}  
% \sum_{\substack{\text{unlabeled trees} \\ \text{with m edges}}} \frac{1}{|\text{Aut(T)}|}  {\cal V}_1 \dots {\cal V}_{m+1}
% \end{split}
% \end{equation}

To evaluate  $\langle O_k \bar O_k\rangle$ on $\mathbb{R}^4$ for $k>2$ we need to run the Gram-Schmidt procedure (\ref{def:GS}). Let us consider the first non-trivial case.
\begin{equation}
O_4 = \Omega_4 - \frac{C_{4,2}}{C_{2,2}} \Omega_2
\end{equation}
thus in order to compute $\langle O_4 \bar O_4 \rangle$ we require
\begin{equation}
\langle O_4 \bar O_4 \rangle = \langle \Omega_4 \Omega_4 \rangle -  \frac{C_{4,2}^2}{C_{2,2}} 
\end{equation}
The fact that $\langle \Omega_2\Omega_2\rangle$ appears in the denominator complicates the task of finding a closed expression for $\langle O_4 \bar O_4 \rangle $. For concreteness, we will limit ourselves to present all the terms with a single value of  the $\zeta$ function. Collecting all such terms we deduce
 %\begin{equation}
  %\langle O_4 \bar O_4 \rangle = \frac{4\lambda^4}{(4\pi)^8} \left( 1+\sum_{n=2}^\infty \zeta(2n-1) \left(\frac{-\lambda}{16 \pi^2}\right)^n 
  %\left( 8 \frac{(2n)!}{n! n!}-\frac{4  (2n)! (2n+2)! (n^2+n+18)}{n! n! (n+1)! (n+4)!}\right)+\dots \right)
%\end{equation} 
\be 
\langle O_4 \bar O_4\rangle = 4\left(\frac{\lambda}{16\pi^2}\right)^4 \left(1-8 \sum_{n=2}^\infty \frac{\zeta_{2n-1}}{n} \left(\frac{-\lambda}{16\pi^2}\right)^n {2n \choose n} \left[ {2n \choose n+4}+{2n \choose n+1}-n \right]+\dots \right)
\label{o4o4r4}
\ee
We can repeat the same procedure for $\langle O_6 \bar O_6\rangle$. A longer computation yields
\be 
\langle O_6 \bar O_6\rangle = 6\left(\frac{\lambda}{16\pi^2}\right)^6 \left(1-12 \sum_{n=2}^\infty \frac{\zeta_{2n-1}}{n} \left(\frac{-\lambda}{16\pi^2}\right)^n {2n \choose n} \left[ {2n \choose n+6}+{2n \choose n+1}-n \right]+\dots \right)
\label{o6o6r4}
\ee

As a first test, these expressions reproduce the terms with a single value of $\zeta$ in \cite{Galvagno:2020cgq}. While we are not writing them down, one can also check that the first terms with a product of two $\zeta$ also agree with the result of \cite{Galvagno:2020cgq}. Now, looking at the explicit expressions (\ref{o2o2r4}, \ref{o4o4r4}, \ref{o6o6r4}) a pattern appears to emerge, so we are led to put forward the following conjecture for generic even $k$,
\be
\langle O_k \bar O_k\rangle  \stackrel{?}{=}  k\left(\frac{\lambda}{16\pi^2}\right)^k \left(1-2k \sum_{n=2}^\infty \frac{\zeta_{2n-1}}{n} \left(\frac{-\lambda}{16\pi^2}\right)^n {2n \choose n} \left[ {2n \choose n+k}+{2n \choose n+1}-n \right]+\dots \right)
\label{twopointconj}
\ee
where the dots stand for terms with two or more values of $\zeta$. Using the Mathematica notebook available in \cite{Galvagno:2020cgq}, we have checked that this conjecture reproduces the first terms of $\langle O_8 \bar O_8\rangle$. As for $\vev{O_k \bar O_k}$ for odd $k$,  a bit of trial and error with the results available in \cite{Galvagno:2020cgq} leads to the following generalized  conjecture
\be
\langle O_k \bar O_k\rangle  \stackrel{?}{=}  k\left(\frac{\lambda}{16\pi^2}\right)^k \left(1-2k \sum_{n=2}^\infty \frac{\zeta_{2n-1}}{n} \left(\frac{-\lambda}{16\pi^2}\right)^n {2n \choose n} \left[ (-1)^k {2n \choose n+k}+{2n \choose n+1}-n \right]+\dots \right)
\label{twopointconj2}
\ee

The conjecture (\ref{twopointconj2}) is appealingly simple. The terms in square brackets has a k-dependent contribution, that is non-vanishing only at orders $n\geq k$, plus a universal, k-independent, contribution. Furthermore, the factor 
\begin{equation*}
\frac{\zeta_{2n-1}}{n} \left(\frac{-\lambda}{16\pi^2}\right)^n {2n \choose n} 
\end{equation*}
coincides with the planar value of certain Feynman diagrams identified in section 7.3 of \cite{Beccaria:2020hgy} (see also \cite{Galvagno:2020cgq}). These ingredients hint at a diagrammatic derivation of (\ref{twopointconj2}). Finally, let us mention that \cite{Beccaria:2020hgy,Beccaria:2021hvt} have developed very efficient techniques to obtain analytic results for certain ${\cal N}=2$ SCFTs, which currently don't include ${\cal N}=2$ SCQD. In these works, a crucial role is played by an infinite matrix that can be written as an integral over Bessel functions. It is straightforward to check that our conjecture (\ref{twopointconj2}) can be written similarly,
\be
\begin{split}
&\sum_{n=2}^\infty \frac{\zeta_{2n-1}}{n} \left(\frac{-\lambda}{16\pi^2}\right)^n {2n \choose n} \left[(-1)^k {2n \choose n+k} +{2n \choose n+1}-n\right]= \\
& \int_0^\infty dw \frac{J_k(\frac{w\sqrt{\lambda}}{\pi})^2-J_1(\frac{w\sqrt{\lambda}}{\pi})^2+\frac{w\sqrt{\lambda}}{2\pi}J_1(\frac{w\sqrt{\lambda}}{\pi})}{2w \sinh^2 w}
\end{split}
\ee
It would be interesting to extend the techniques of \cite{Beccaria:2020hgy,Beccaria:2021hvt} to arbitrary ${\cal N}=2$ Lagrangian SCFTs; this would allow to prove (\ref{twopointconj2}) and extend it to terms with two or more values of $\zeta$.

Let's now switch to the determination of planar 3-point functions on $\bR^4$, repeating the same procedure. The first non trivial extremal $3-$point function is given by
\begin{equation}
\langle O_2 O_2 \bar O_4 \rangle = \langle \Omega_2 \Omega_2 \Omega_4 \rangle - \frac{C_{4,2}}{C_{2,2}} \langle \Omega_2 \Omega_2 \Omega_2 \rangle \, ,
\end{equation}
Upon collecting all the terms with a single $\zeta$ we obtain

%\begin{equation}
%\begin{split}
%\langle O_2 O_2 \bar O_4 \rangle = -\frac{16\lambda^4}{(4\pi)^8} \Biggl( 1 - & 2\sum \frac{\zeta(2i+2j-1)}{(i+j)} \left(\frac{-\lambda}{16\pi^2}\right)^{i+j} {2i+2j\choose 2i} \frac{(2i)!(2j)!}{(i-1)!i!(j+1)!j!}  \\
%&  \left( \frac{3(i+j+1)(i+j+2)}{(i+2)}-\frac{(i+j)^2(i+j+1)}{(i+1)i} \right) \Biggr)
%\end{split}
%\end{equation}

\begin{equation}
\langle O_2 O_2 \bar O_4 \rangle = \textnormal{N}^{-1}\frac{16\lambda^4}{(4\pi)^8} \Biggl( 1 -  2\sum_{n=2}^\infty \zeta_{2n-1} \left(\frac{-\lambda}{16\pi^2}\right)^n {2n \choose n} \left( \frac{(2n+1)! (n^2+3n+12)}{(n+3)! n!}-(n+3) \right) \Biggr)+\dots
\end{equation}
To get rid of the ambiguity associated to the normalization of the CPOs, it is convenient to define the normalized 3-point functions,
\begin{equation}
\langle O_{k_1} O_{k_2} \bar{O}_{k1+k2} \rangle_n= \frac{\langle O_{k_1} O_{k_2} \bar{O}_{k1+k2} \rangle}{\textnormal{N} \sqrt{\langle O_{k_1} \bar O_{k_1}\rangle \langle O_{k_2} \bar O_{k_2}\rangle \langle
O_{k_1+k_2} \bar{O}_{k1+k2} \rangle}}
\end{equation}
After doing so, we find
%\begin{equation}
%\langle O_2 O_2 \bar O_4 \rangle_n =4-8\sum_{n=2}^\infty \zeta(2n-1) \left(\frac{-\lambda}{16\pi^2}\right)^n {2n \choose n} (n-1)
%\left(  \frac{(n^2+4n+12)(2n+1)!}{n!(n+4)!}-1\right)+\dots
%\end{equation}

\begin{equation}
\frac{\langle O_2 O_2 \bar O_4 \rangle_n}{\sqrt{2\cdot 2\cdot4}} = \textnormal{N}^{-1}\left[1-\sum_{n=2}^\infty \zeta_{2n-1} \left(\frac{-\lambda}{16\pi^2}\right)^n {2n \choose n}
\left[{2n \choose n+2}+{2n\choose n+2}+{2n \choose n+4}+(n-1)(C_n-2)\right]+\dots \right]
\end{equation}
Repeating all these steps for $\langle O_2 O_4 \bar O_6 \rangle_n$ we find
\begin{equation}
\frac{\langle O_2 O_4 \bar O_6 \rangle_n}{\sqrt{2\cdot 4\cdot 6}} = \textnormal{N}^{-1}\left[1-\sum_{n=2}^\infty \zeta_{2n-1} \left(\frac{-\lambda}{16\pi^2}\right)^n {2n \choose n}
\left[{2n \choose n+2}+{2n\choose n+4}+{2n \choose n+6}+(n-1)(C_n-2)\right]+\dots \right]
\end{equation}
The first terms of these expressions reproduce the results presented in \cite{Baggio:2016skg}. These two computations suggest the following general conjecture for planar 3-point functions of even-dimensional operators
\begin{equation}
\begin{split}
\frac{\langle O_{k_1} O_{k_2} \bar{O}_{k1+k2} \rangle_n}{\sqrt{k_1 \cdot k_2 \cdot (k_1+k_2)}}  \stackrel{?}{=} \textnormal{N}^{-1}&\Biggl[ 1-\sum_{n=2}^\infty \left(\frac{-\lambda}{16\pi^2}\right)^n \zeta_{2n-1} {2n \choose n} \\
&\left( {2n \choose n+k_1} + {2n \choose n + k_2}+{2n \choose n+k_1+k_2}+(n-1) (\mathcal{C}_n-2) \right) \Biggr]+\dots
\end{split}
\label{threepointconj}
\end{equation}

%\begin{equation}
%\begin{split}
%\frac{\langle O_{k_1} O_{k_2} \bar{O}_{k1+k2} \rangle_n}{\sqrt{k_1 \cdot k_2 \cdot (k_1+k_2)}} = & 1-\sum_{n=2}^\infty \left(\frac{-\lambda}{16\pi^2}\right)^n \zeta_{2n-1} {2n \choose n} \\
%&\left[ {2n \choose n+k_1} + {2n \choose n + k_2}+{2n \choose n+k_1+k_2}+(n-1) (\mathcal{C}_n-2) \right] +\dots
%\end{split}
%\end{equation}

%\begin{equation}
%\begin{split}
%\frac{\langle O_{k_1} O_{k_2} \bar{O}_{k1+k2} \rangle_n}{\sqrt{k_1 \cdot k_2 \cdot (k_1+k_2)}}  \stackrel{?}{=}  &  \\
%1-\sum_{n=2}^\infty \left(\frac{-\lambda}{16\pi^2}\right)^n \zeta_{2n-1} {2n \choose n} &\left[ {2n \choose n+k_1} + {2n \choose n + k_2}+{2n \choose n+k_1+k_2}+(n-1) (\mathcal{C}_n-2) \right] +\dots
%\end{split}
%\end{equation}
As a first check,  this conjecture correctly reproduces for arbitrary even $k_1,k_2$ the $\zeta_3$ term found in \cite{Baggio:2016skg}. We have checked that it also correctly reproduces the first terms of $\langle O_4 O_4 \widebar O_8\rangle_n$ and $\langle O_4 O_6 \widebar O_{10}\rangle_n$\footnote{We have computed the frist terms of these 3-point functions explicitly, using the Gram-Schmidt procedure. The results obtained agree with the conjecture (\ref{threepointconj}). However, we do not agree with the coefficient of $\zeta_7$ for $\langle O_4 O_6 \widebar O_{10}\rangle_n$ presented in \cite{Baggio:2016skg}.}. Again, we find the form of this conjecture remarkably simple, and suspect that it hints at the existence of a direct derivation of these results, that bypasses going through the route of computing first correlators on $S^4$. Finally, it is also possible to have a non-vanishing 3-point function involving two odd and one even operators. Motivated by (\ref{twopointconj2}) a possible guess is that in this case the ${2n \choose n+k}$ factors in (\ref{threepointconj}) pick up a minus sign for odd $k$, but we haven't checked explicitly.

%\begin{equation}
%\begin{split}
%\frac{\langle O_2 O_4 \bar{O}_6 \rangle_n}{\sqrt{2\cdot 4\cdot 6}} &= \Biggl(1-2 \sum_{ij} \frac{c_{2i2j}(2i)!(2j)!}{i!(i-1)!(j!)(j-1)!} \left(\frac{\lambda}{16\pi^2} \right)^2\frac{1}{i (i+1) (i+2) (i+3) j (j+1) (j+2) (j+3)} \\
%& \Biggl(4i^5(6+5j+j^2)+i^4(252+228j+43j^2+5j^3)+i^3(912+490j-228j^2+4j^3-2j^4) \\
%&-6j(114+227j + 147j^2+37j^3+3j^4)+i^2(1332+984j+13j^2+276j^3-34j^4-3j^5) \\
%&-3i(-216+10j+322j^2+149j^3+66j^4+5j^5) \Biggr)\Biggr)
%\end{split}
%\end{equation}

\acknowledgments
We would like to thank Matteo Beccaria, Francesco Galvagno, Vassilis Niarchos and Kyriakos Papadodimas for correspondence. We would also like to thank Marcos Mari\~no for sharing a Mathematica notebook written for \cite{Grassi:2014vwa}, that allowed us to check our results in section 2. We would also like to thank Francesco Galvagno and Michelangelo Preti for making publicly available the Mathematica notebook of  \cite{Galvagno:2020cgq}. Research supported by  the State Agency for Research of the Spanish Ministry of Science and Innovation through the ``Unit of Excellence Mar\'ia de Maeztu 2020-2023'' award to the Institute of Cosmos Sciences (CEX2019-000918-M) and PID2019-105614GB-C22, and by AGAUR, grant 2017-SGR 754.  A. R. F. is further supported by an FPI-MINECO fellowship. 

 \appendix

\section{Explicit expansions}

In the main text, when writing explicit results, we have mostly restricted to displaying only terms with a single value of the $\zeta$ function. Our techniques can equally well produce terms with products of $\zeta$s. In this appendix we present two examples, the planar limit of $\langle O_2 \widebar{O_2} \rangle$ and $\langle O_2 O_2 \bar{O}_4 \rangle$ on $\bR^4$.
\begin{equation}
\begin{split}
\langle O_2 \widebar{O_2} \rangle = \frac{2\lambda^2}{(4\pi)^4}&\Biggl[1-\frac{9}{4}\zeta_3\frac{\lambda^2}{(2\pi)^4} +\frac{15}{2}\zeta_5\frac{\lambda^3}{(2\pi)^6} +\frac{5}{8}\left(9 \zeta _3^2-35 \zeta _7\right)\frac{\lambda^4}{(2\pi)^8} -\frac{45}{64}\left(60 \zeta _3 \zeta _5-91 \zeta _9\right) \frac{\lambda^5}{(2\pi)^{10}} \\ 
&+\frac{21}{512}\left(-360 \zeta _3^3+3360 \zeta_7 \zeta _3+1900 \zeta _5^2-4697 \zeta _{11}\right)\frac{\lambda^6}{(2\pi)^{12}} \\
&+\frac{7}{256}{\left(20 \left(\zeta _5 \left(324 \zeta _3^2-917 \zeta _7\right)-819 \zeta _3 \zeta _9\right)+21879 \zeta _{13}\right)}\frac{\lambda^7}{(2\pi)^{14}} \\
&+\frac{27}{4096}\Bigg( 6048 \zeta _3^4-94080 \zeta _7 \zeta _3^2+528 \left(427 \zeta _{11}-200 \zeta _5^2\right) \zeta _3 \\
&+140 \left(861 \zeta _7^2+1744 \zeta _5 \zeta _9\right)-289575 \zeta _{15}\Bigg)\frac{\lambda^8}{ (2\pi)^{16}} \\
&-\frac{15}{16384}\Bigg(560 \left(\zeta _5 \left(1296 \zeta _3^3-9333 \zeta _7 \zeta _3-1750 \zeta _5^2\right)+9 \left(1091 \zeta _7-468 \zeta _3^2\right) \zeta _9\right) \\
&+5775660 \zeta _5 \zeta _{11}+5513508 \zeta _3 \zeta _{13}-6804369 \zeta _{17}\Bigg) \frac{\lambda^9}{(2\pi)^{18}}  \\
& -\frac{11}{32768}\Bigg(326592 \zeta _3^5-7257600 \zeta _7 \zeta _3^3+4860 \left(4697 \zeta _{11}-2500 \zeta _5^2\right) \zeta _3^2 \\
&+300 \left(80164 \zeta _7^2+162876 \zeta _5 \zeta _9-173745 \zeta _{15}\right) \zeta _3-23481360 \zeta _9^2\\
&+525 \zeta _7 \left(51660 \zeta _5^2-92851 \zeta _{11}\right)-53088750 \zeta _5 \zeta _{13}+61708504 \zeta _{19}\Bigg)\frac{\lambda^{10}}{ (2\pi)^{20}}\Biggr]
\end{split}
\end{equation}

\begin{equation}
\begin{split}
\langle O_2 O_2 \bar{O}_4 \rangle = 4\textnormal{N}^{-1} &\Biggl( 1 - \frac{3 \zeta_3 \lambda^2}{64\pi^4}+\frac{45 \zeta_5 \lambda^3}{512\pi^6}+ \frac{3(72 \zeta_3^2-1085 \zeta_7)}{32768\pi^8}\lambda^4 + \frac{45(287\zeta_9-64 \zeta_3 \zeta_5)}{131072\pi^{10}}\lambda^5 \\
&+\frac{3 \left(16164 \zeta _3^3+19075 \zeta _7 \zeta _3+10500 \zeta _5^2-65681 \zeta _{11}\right) \lambda ^6}{2097152 \pi ^{12}} \\
&-\frac{15 \left(\zeta _5 \left(80550 \zeta _3^2+36911 \zeta _7\right)+34272 \zeta _3 \zeta _9-99099 \zeta _{13}\right) \lambda ^7}{16777216 \pi
   ^{14}}\\
&+\Bigl(3822336 \zeta _3^4+47231184 \zeta _7 \zeta _3^2+32 \left(1633325 \zeta _5^2+738969 \zeta _{11}\right) \zeta _3 \\
&+245 \left(48493 \zeta_7^2+100032 \zeta _5 \zeta _9-243672 \zeta _{15}\right)\Bigr) \frac{3\lambda^8}{2147483648 \pi ^{16}} + \cdots \Biggr)
\end{split}
\end{equation}

\bibliographystyle{JHEP}

\end{document}